\documentclass[prl,aps,twocolumn,floatfix]{revtex4}
\usepackage{graphicx,graphics,psfrag,amsmath,calc}
\usepackage{epsfig}
\usepackage{color,comment}
\begin{document}

\title{Unconventional color superfluidity in ultra-cold fermions: \\
Quintuplet pairing, quintuple point and pentacriticality}

\author{
Doga Murat Kurkcuoglu and C. A. R. S{\'a} de Melo
}

\affiliation{
School of Physics, Georgia Institute of Technology, Atlanta, 
Georgia 30332, USA
}

\date{\today}

\begin{abstract}
We describe the emergence of color superfluidity in 
ultra-cold fermions induced by color-orbit and color-flip fields 
that transform a conventional singlet-pairing s-wave system into 
an unconventional non-s-wave superfluid with quintuplet pairing. 
We show that the tuning of interactions, color-orbit and color-flip
fields transforms a momentum-independent scalar order 
parameter into an explicitly momentum-dependent tensor order parameter. 
We classify all unconventional superfluid phases 
in terms of the {\it loci} of zeros of their quasi-particle 
excitation spectrum in momentum space and we identify several 
Lifshitz-type topological transitions.  Furthermore, when boundaries 
between phases are crossed, non-analyticities in the compressibility arise.
We find a quintuple point, which is also pentacritical, where four gapless 
superfluid phases converge into a fully gapped superfluid phase.

\end{abstract}
\maketitle

%
%

Ultra-cold fermions with two internal states trapped in harmonic, 
box and optical lattice potentials can serve as simulators of models often found 
in condensed matter physics. Key examples are experimental simulations of 
the crossover from Bardeen-Cooper-Schrieffer (BCS) 
to Bose-Einstein Condensation (BEC) superfluidity
~\cite{jin-2003, hulet-2003, ketterle-2003, jin-2004a, 
salomon-2004, grimm-2004,thomas-2004, ketterle-2005}, 
and more recently, experimental simulations 
of Fermi-Hubbard systems in optical lattices~\cite{hulet-2015, greiner-2017}, 
where an antiferromagnetic insulator at half-filing is expected to become 
a d-wave superfluid upon hole doping~\cite{bickers-1989}.

Furthermore, ultra-cold fermions 
with three internal states can provide 
insights~\cite{zhuang-2006, hofstetter-2007, demler-2007, sademelo-2008, 
baym-2010, nishida-2012, kurkcuoglu-2015, kurkcuoglu-2018a} 
into models of quantum 
chromodynamics (QCD), where baryon formation (bound trimers) at nuclear 
density quark-matter and color superconductivity 
is expected to occur at densities compatible with those of 
neutron stars~\cite{alford-2008}. Candidates for the 
experimental studies of color superfluidity are three-component mixtures
of atomic fermions, such as, $^6$Li, $^{40}$K or $^{173}$Yb. 
Some experimental work on the collisional stability of three component
mixtures of $^6$Li exists~\cite{jochim-2008, ohara-2009}, and 
mixtures of up to six components in $^{173}$Yb have 
been reported~\cite{takahashi-2009}, where SU(6)-symmetric fermion phases 
have been shown to exist~\cite{takahashi-2010, takahashi-2012}.

Color-orbit coupling and color-flip fields can 
be implemented in ultra-cold fermions using the Raman scheme to 
create spin-orbit and Rabi couplings in spin-1 bosons~\cite{spielman-2015}.
This has been achieved also for spin-1/2 
fermions~\cite{spielman-2013, zhang-2014}, where interactions were tuned,
but temperatures remained high~\cite{supplementary-material}. 
Novel methods of reducing the temperature and of creating artificial 
spin-orbit or color-orbit fields in the laboratory are being 
sought~\cite{spielman-2018} using radio-frequency chip 
technology~\cite{spielman-2010}.  Color superfluidity in the context 
of ultra-cold fermions is different from SU(3)-symmetric 
color superconductivity in QCD, because many parameters in the 
Hamiltonian can be tuned, such as interactions in $^6$Li and $^{40}$K 
via magnetic Fano-Feshbach resonances~\cite{chin-2010} or in $^{173}$Yb via
orbital Fano-Feshbach resonances~\cite{fallani-2015, bloch-2015}. 
In this paper, we take advantage of the existing tunability in ultra-cold 
fermions to propose the existence of unconventional color 
superfluids with quintuplet pairing in the presence of color-orbit coupling.
When s-wave interactions and color-flip fields 
are changed, the resulting phase diagram is very rich, containing 
a quintuple and pentacritical point, where the compressibility is 
non-analytic and four gapless superfluid phases converge into 
a fully gapped one.

%
%

To describe interacting three-color fermions under the influence of 
color-orbit and color-flip fields, 
we start with a general independent-particle Hamiltonian that results 
from the coupling to a spatially modulated chip or 
Raman beams~\cite{kurkcuoglu-2015, kurkcuoglu-2018a}
\begin{equation}
\label{eqn:color-hamiltonian}
{\bf H}_0({\bf k}) 
= 
\varepsilon ({\bf k}){\bf 1} - 
h_x({\bf k}){\bf J}_x - 
h_z({\bf k}){\bf J}_z + 
b_z {\bf J}_z^2 ,
\end{equation}
where ${\bf J}_{\ell}$ are spin-one angular momentum matrices
with $\ell = \{ x, y, z \}$.  
The reference kinetic energy
$
\varepsilon ({\bf k}) = {\bf k}^2/(2m) + \eta
$ 
is the same for all colors,
$
h_x({\bf k}) = -\sqrt{2} \Omega
$ 
is the color-flip Rabi field, and 
$
h_z({\bf k}) = 2k_T k_x /(2m) + \delta
$
is a momentum dependent Zeeman field along the $z$-axis, 
which is transverse to the momentum transfer direction along the
$x$-axis, 
and
$
b_z = k_T^2/(2m) - \eta
$
is the quadratic color-shift term. 
Notice that $h_z ({\bf k})$ contains
the color-orbit coupling term $2k_T k_x /(2m)$ 
as well as a color-shift term controlled by the detuning $\delta$. 
A similar hamiltonian was studied for spin-one 
bosons~\cite{ohberg-2014, spielman-2015}.

The chip-atom or Raman-atom interaction Hamiltonian 
can be written in second-quantized notation as 
\begin{equation}
\label{eqn:independent-particle-hamiltonian-operator}
{H}_{{\rm CA}}
= 
\sum_{\bf k}
{\bf \Psi}^{\dagger}_{\bf k}
{\bf H}_0({\bf k})
{\bf \Psi}_{\bf k}
\end{equation}
where the spinor creation operator is 
$
{\bf \Psi}^{\dagger}_{\bf k} 
= 
\left[
\psi_R^{\dagger}({\bf k}),
\psi_G^{\dagger}({\bf k}),
\psi_B^{\dagger}({\bf k})
\right]
$,
with
$
\psi_c^{\dagger}({\bf k})
$
creating a fermion label by momentum ${\bf k}$ in 
color state $c = \{ R, G, B \}$~\cite{supplementary-material}. 
We use as units 
the Fermi energy 
$
E_F = k_F^2/(2m)
$
and the Fermi momentum 
$
k_F = (2 \pi^2 n)^{1/3},
$
based on the total density 
of fermions $n = 3 k_F^3/(6 \pi^2)$ with initial identical
kinetic energies $\epsilon_{\bf k} = {\bf k}^2/(2m)$ 
for all three internal states. 
This means that our reference system is that with all parameters
$\eta, k_T, \Omega$ and $\delta$ set to zero. 

\begin{figure} [tb]
\includegraphics[width = 0.49\linewidth]{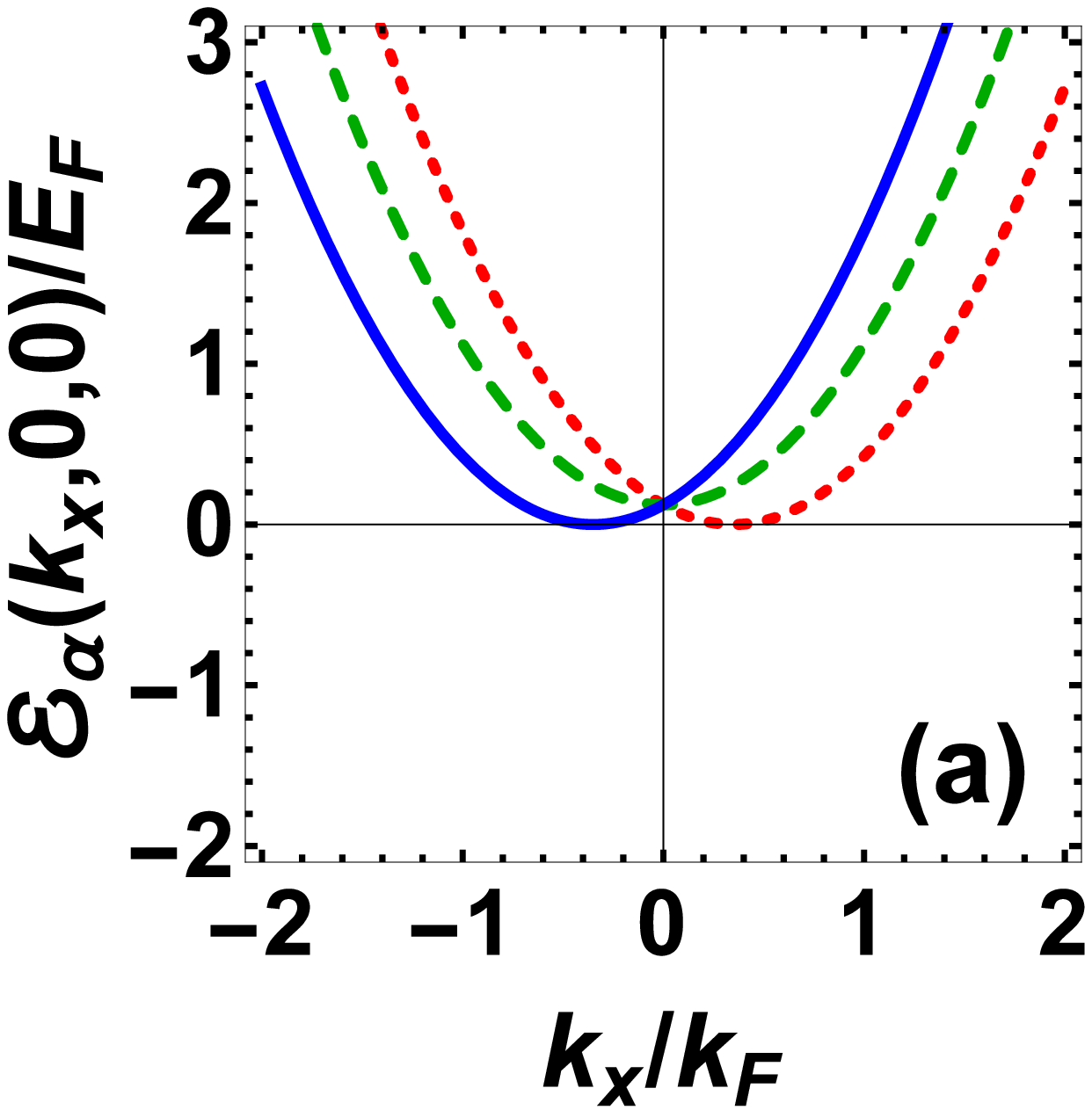}
\includegraphics[width = 0.49\linewidth]{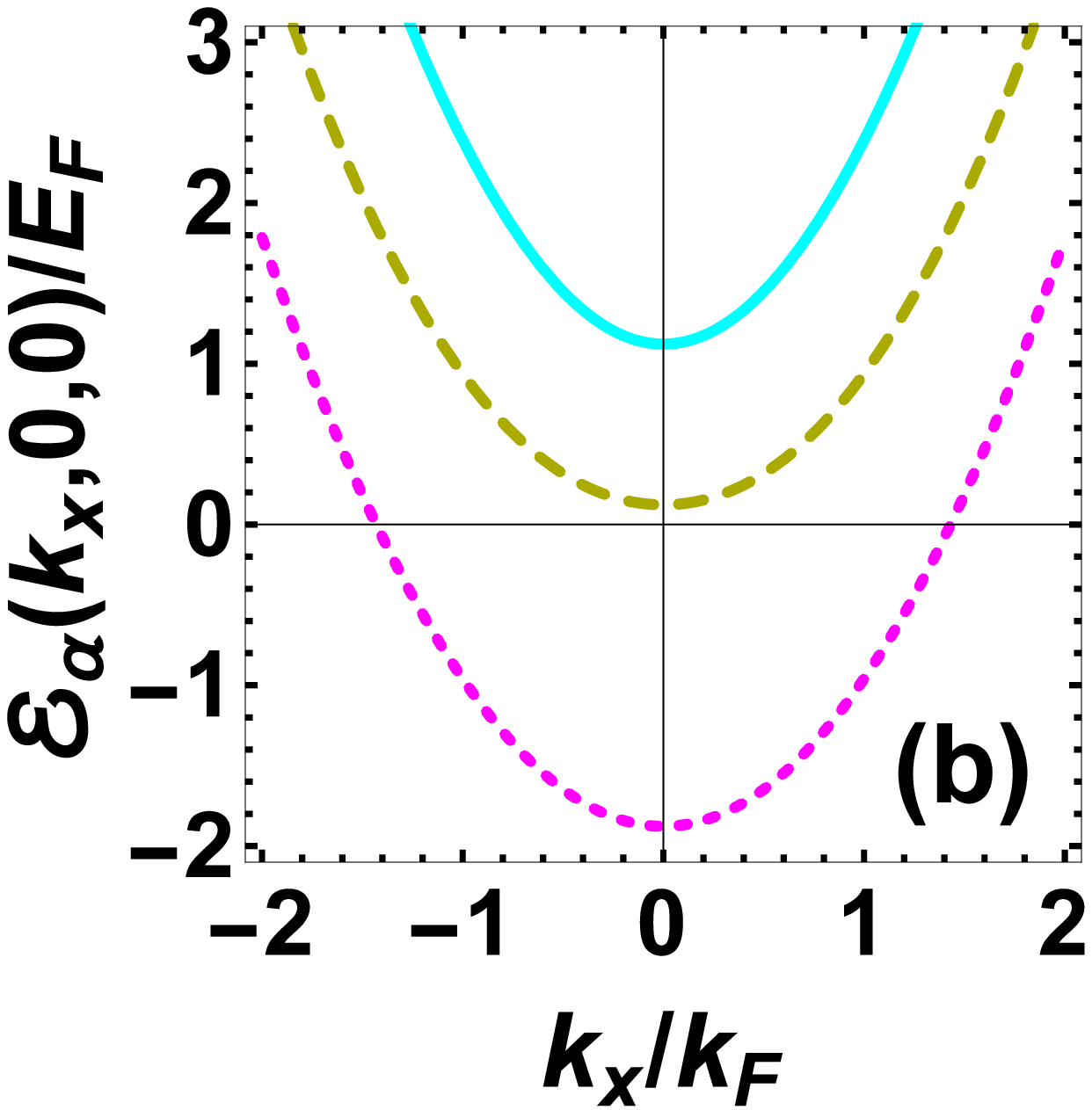}
\caption{
\label{fig:one}  
(Color Online) 
Eigenvalues $\mathcal{E}_{\alpha} ({\bf k})$ versus $k_x$ 
for momentum transfer $k_T = 0.35k_F$ 
and quadratic color shift $b_z = 0$, that is, $\eta = k_T^2/2m$. 
The cases with color-flip (Rabi) frequencies $\Omega = 0$ and
$\Omega = E_F$ are shown in a) and b), respectively. 
In a) states $\{R, G, B\}$ are not mixed:
the dotted-red line corresponds to $\mathcal{E}_{\Uparrow} ({\bf k})$,
the dashed-green line to $\mathcal {E}_0 ({\bf k})$, and
the solid-blue line to $\mathcal{E}_{\Downarrow} ({\bf k})$.
In b) states $\{R, G, B\}$ are mixed:
the dotted-magenta line corresponds 
to $\mathcal{E}_{\Uparrow} ({\bf k})$,
the dashed-yellow line to $\mathcal {E}_0 ({\bf k})$, and
the solid-cyan line to $\mathcal{E}_{\Downarrow} ({\bf k})$.
}
\end{figure}

In Fig.~{\ref{fig:one}}, we show  
eigenvalues $\mathcal{E}_{\alpha} ({\bf k})$ of ${\bf H}_0 ({\bf k})$ 
versus momentum $k_x$ for fixed momentum transfer $k_T = 0.35 k_F$
and zero detuning $\delta = 0$~\cite{supplementary-material}. 
In the chip setup, the momentum transfer is
$k_T = 2\pi/\lambda_C$, where $\lambda_C$ reflects the chip's spatial 
modulation. A characteristic value is
$\lambda_C = 2\mu m$, leading to a typical fermion density
$
n = (4 \pi/\lambda_C^3)(k_F / k_T)^3
= 3.6 \times 10^{13} {\rm cm}^{-3}
$ 
for $k_T = 0.35 k_F$.
We choose the value $k_T = 0.35 k_F$
such that all qualitative changes in superfluid phases occur in the
interval $-2 < 1/(k_F a_s) < 2$ for fixed $\Omega$ as
discussed later. In Fig.~{\ref{fig:one}a}, the Rabi frequency 
is $\Omega = 0$ and the states $\{R, G, B \}$ remain 
an eigenbasis, and thus are uncoupled. In this case, along 
the $k_x$ direction, the energy dispersion of state $R$$(B)$ 
is shifted to the right (left) by $k_T$$(-k_T)$, and the energy
dispersion of state $G$ is unshifted. However, for $\Omega \ne 0$ 
these states mix and are no longer eigenstates of the independent 
particle Hamitonian defined in 
Eqs.~(\ref{eqn:color-hamiltonian}) and 
~(\ref{eqn:independent-particle-hamiltonian-operator}) 
with $b_z = 0$. One of the effects of $\Omega$ is to lift
degeneracies, but when $\Omega$ is large $(\Omega = E_F)$, 
as shown in Fig.~\ref{fig:one}b, the states 
$\{R, G, B \}$ are strongly mixed, and the 
eigenergies are largely separated from each other.

%

To study the quantum phases of three-color fermions, we add
interactions between atoms in different internal states and consider
contact attractive interactions 
$-g_{cc^{\prime}} \delta ({\bf r} - {\bf r}^{\prime})$ of 
strength $g_{c c^{\prime}} > 0$, between internal 
states $c \ne c^{\prime}$ only. 
The atom-atom 
interaction Hamiltonian can be written in momentum space
as
\begin{eqnarray}
\label{eqn:atom-atom-interaction-hamiltonian}
H_{{\rm AA}} 
= 
-
\frac{1}{V} 
\sum \limits_{{\bf Q},\left\{c \neq c^{\prime}\right\}} 
g_{cc^{\prime}} 
b_{cc^{\prime}}^{\dagger}({\bf Q})
b_{cc^{\prime}}({\bf Q}),
\end{eqnarray}
where 
$V$ is the volume, ${\bf Q}$ 
is the center-of-mass momentum of fermion pairs characterized
by the operator 
$
b_{cc^{\prime}}^{\dagger}({\bf Q}) 
= 
\sum_{\bf k} 
\psi_c^{\dagger}({\bf k} + {\bf Q}/2)
\psi_{c^{\prime}}^{\dagger}(-{\bf k} - {\bf Q}/2).
$ 
Thus, the Hamiltonian describing the effects of color-orbit coupling,
color-flip fields and atom-atom interactions is 
\begin{equation}
H
=
H_{{\rm CA}}
+
H_{{\rm AA}}
-
\mu {\hat N},
\end{equation}
where 
$
{\hat N} 
= 
\sum_{c,{\bf k}}
\psi_c^{\dagger}({\bf k})
\psi_c({\bf k})
$ 
represents the total number of particles.
We focus on uniform superfluid phases with 
${\bf Q} = {\bf 0}$ and order parameter tensor 
$
\Delta_{cc^{\prime}}
= 
-g_{cc^{\prime}}
\langle 
b_{cc^{\prime}}
(
{\bf 0}
)
\rangle/V,
$
leading to the Hamiltonian~\cite{supplementary-material}
\begin{equation}
\label{eqn:mean-field-Hamiltonian}
H_{\rm MF} 
= 
\frac{1}{2}
\sum_{\bf k}
{\bf \Psi}^{\dagger}_{N} ({\bf k})
{\bf H}_{\rm MF}({\bf k})
{\bf \Psi}_{N} ({\bf k})
+
V \sum_{c\neq c^{\prime}}
\frac{|\Delta_{cc^{\prime}}|^2}{g_{cc^{\prime}}}
+
{\mathcal C} (\mu) 
\end{equation}
where the six-component Nambu spinor is 
$
{\bf \Psi}^{\dagger}_{N} ({\bf k})
=
\left[
\Psi_R^{\dagger}({\bf k}),
\Psi_G^{\dagger}({\bf k}),
\Psi_B^{\dagger}({\bf k}),
\Psi_R ({-\bf k}),
\Psi_G ({-\bf k}),
\Psi_B ({-\bf k})
\right].
$
Here, the function
$
{\mathcal C} (\mu) 
= 
\frac{1}{2}
\sum_{{\bf k} c}
\xi_c(-{\bf k})
$
contains the term 
$
\xi_c({\bf k}) 
= 
\varepsilon_c({\bf k}) -\mu
$
representing the residual kinetic energies, 
with 
$
\varepsilon_c({\bf k})
$
representing the diagonal matrix elements 
of ${\bf H}_0 ({\bf k})$ given by
$
\varepsilon_R ({\bf k}) 
= 
\varepsilon ({\bf k})
- 
h_z ({\bf k})
+
b_z;
$
$
\varepsilon_G ({\bf k}) 
= 
\varepsilon ({\bf k});
$
and
$
\varepsilon_B ({\bf k}) 
= 
\varepsilon ({\bf k})
+
h_z ({\bf k})
+
b_z.
$
The 
$
6 \times 6
$
Hamiltonian matrix is
\begin{eqnarray}
\label{eqn:mean-field-hamiltonian-matrix}
{\bf H}_{\rm MF}({\bf k})
=
\left(
\begin{array}{cc}
\overline {\bf H}_0({\bf k})	&	{\bf \Lambda}		\\
{\bf \Lambda}^{\dagger}	&	-\overline {\bf H}_0^* (-{\bf k})
\end{array}
\right),
\end{eqnarray}
where the diagonal block matrix is 
$
\overline {\bf H}_0({\bf k})
=
{\bf H}_0({\bf k})
-\mu {\bf 1}
$
and the off-diagonal block matrix is
\begin{eqnarray}
\label{eqn:order-parameter-tensor}
{\bf \Lambda}
=
\left(
\begin{array}{ccc}
0		&	\Delta_{RG}	&	\Delta_{RB}	\\
-\Delta_{RG}	&	0		&	\Delta_{GB}	\\
-\Delta_{RB}	&	-\Delta_{GB}	& 	0
\end{array}
\right),
\end{eqnarray}
representing the order parameter tensor. 
In this work, we consider the simpler case 
where $g_{RG} = g_{GB} = 0$, and $g_{RG} = g$, 
which leads to $\Delta_{RG} = \Delta_{GB} = 0$,
and $\Delta_{RB} = \Delta$, such that the order parameter 
tensor $\Delta_{cc^\prime}$ is characterized by a single 
complex scalar $\Delta$. For example, this choice reflects 
the experimental condition of three internal states of 
trapped $^{40}$K in the vicinity of its s-wave Fano-Feshbach resonance 
near $200$~Gauss, where states 
$\vert R \rangle = \vert 9/2, -9/2 \rangle$ and 
$\vert B \rangle = \vert 9/2, -7/2 \rangle$ interact, but state
$\vert G \rangle = \vert 9/2, -5/2 \rangle$ does not interact 
with any other state~\cite{jin-2004b}. This situation corresponds 
to a single-channel pairing color superfluid, where only 
$R$ and $B$ fermions experience attractive s-wave interactions.

The corresponding thermodynamic potential is~\cite{supplementary-material} 
\begin{equation}
\label{eqn:thermodynamic-potential}
{\mathcal Q}_{\rm MF}
=
-\frac{T}{2}
\sum_{{\bf k} j} 
\ln 
\left\{
1 
+
\exp \left[ - \frac{E_j ({\bf k})}{T} \right] 
\right\}
+
V \frac{\vert \Delta_{RB} \vert^2}{g_{RB}}
+
{\mathcal C} (\mu)
\end{equation}
where $j = \{ 1,\cdots,6 \}$ labels eigenergies $E_j ({\bf k})$
of ${\bf H}_{\rm MF} ({\bf k})$ in 
Eq.~(\ref{eqn:mean-field-hamiltonian-matrix}).
Minimizing ${\mathcal Q}_{\rm MF}$ with respect to $\Delta_{RB}^*$ 
via $\delta {\mathcal Q}_{\rm MF}/\delta \Delta_{RB}^* = 0$
leads to the order parameter equation
\begin{equation}
\label{eqn:order-parameter-equation}
\frac{V}{g_{RB}} \Delta_{RB}
=
\frac{1}{2} 
\sum \limits_{\bf k}
\sum \limits_{j=1}^3
\tanh\left(
\frac{E_j({\bf k})}{2 T}
\right) 
\frac{\partial E_j({\bf k})}{\partial \Delta_{RB}^*},
\end{equation}
and fixing the total number of particles via 
$N = - \partial{\mathcal Q}_{\rm MF}/\partial \mu \vert_{T,V}$
leads to the number equation
\begin{equation}
\label{eqn:number-equation}
N 
=  
\frac{1}{2}
\sum \limits_{\bf k}
\left[
\sum \limits_{j=1}^3
\tanh\left(
\frac{E_j({\bf k})}{2T}
\right) 
\frac{\partial E_j({\bf k})}{\partial \mu}
+
3
\right].
\end{equation}
The sum over $j$ involves only 
quasiparticle energies $(j = \{1, 2, 3\})$, 
because we used quasiparticle/quasihole 
symmetry to eliminate the quasihole energies. 
Using the relation
$
V/g_{RB} 
= 
-mV/(4 \pi a_s) 
+ 
\sum_{\bf k}1/(2 \epsilon_{\bf k}),
$
we express the 
bare coupling constant $g_{RB}$ in terms of the scattering length $a_s$ 
in the absence of the color-orbit and color-flip fields.
We note also that 
Eqs.~(\ref{eqn:thermodynamic-potential}),~(\ref{eqn:order-parameter-equation}) 
and~(\ref{eqn:number-equation}) are only valid at low temperatures, that is, 
$T \ll E_F$, as they do not include amplitude and phase fluctuations of the
order parameter that become increasingly more important as temperature 
is raised from zero. Thus, in the
remainder of the paper, we choose the 
specific value of $T = 0.02 E_F$ to illustrate the 
low temperature regime of phase diagrams, such that all qualitative 
changes occur in the experimental range $-2 < 1/(k_F a_s) < 2$.

%
\begin{figure} [b]
\centering
\epsfig{file=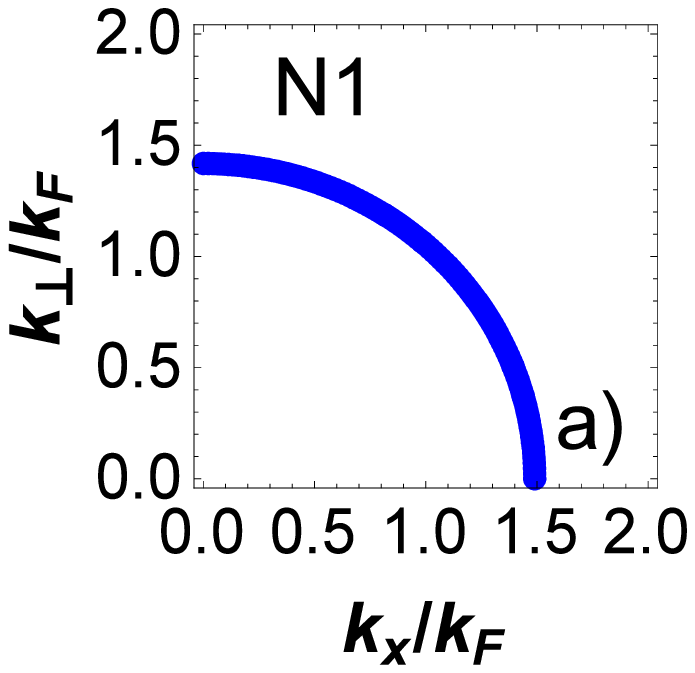,width=0.32 \linewidth}
\epsfig{file=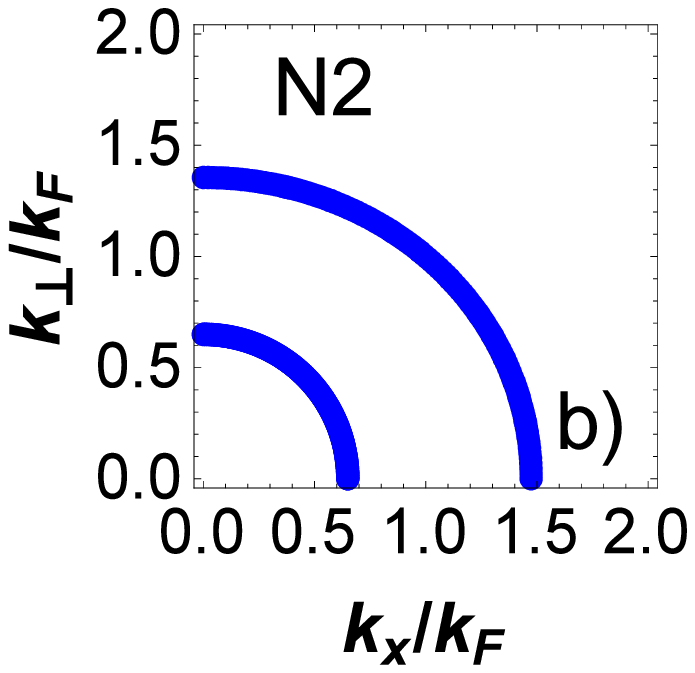,width=0.32 \linewidth}
\epsfig{file=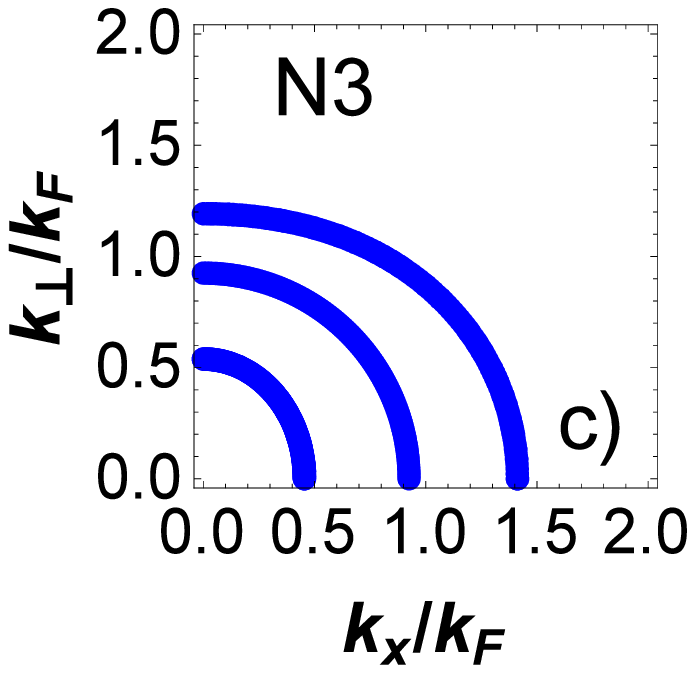,width=0.32 \linewidth}
\\
\epsfig{file=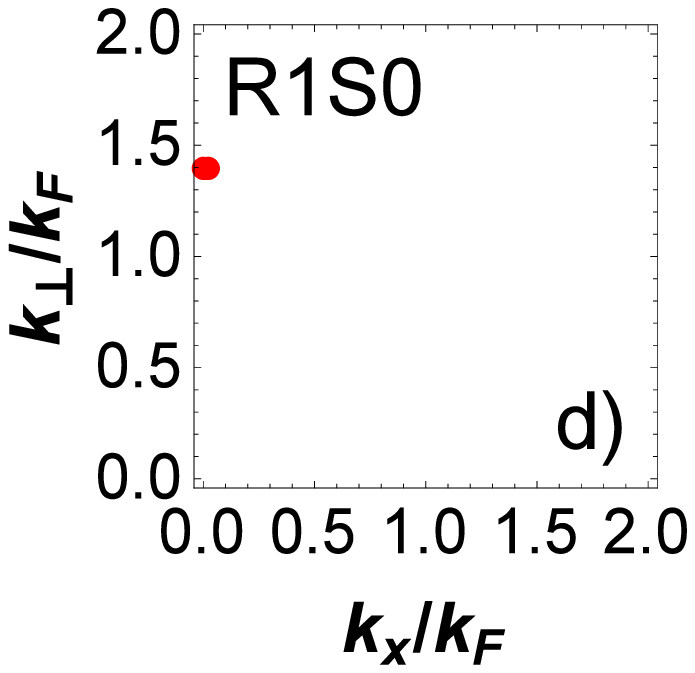,width=0.32 \linewidth}
\epsfig{file=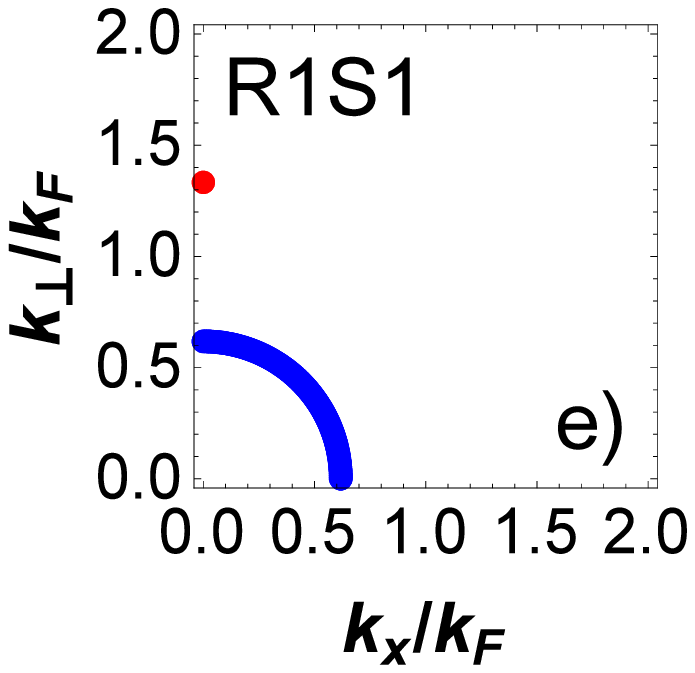,width=0.32 \linewidth}
\epsfig{file=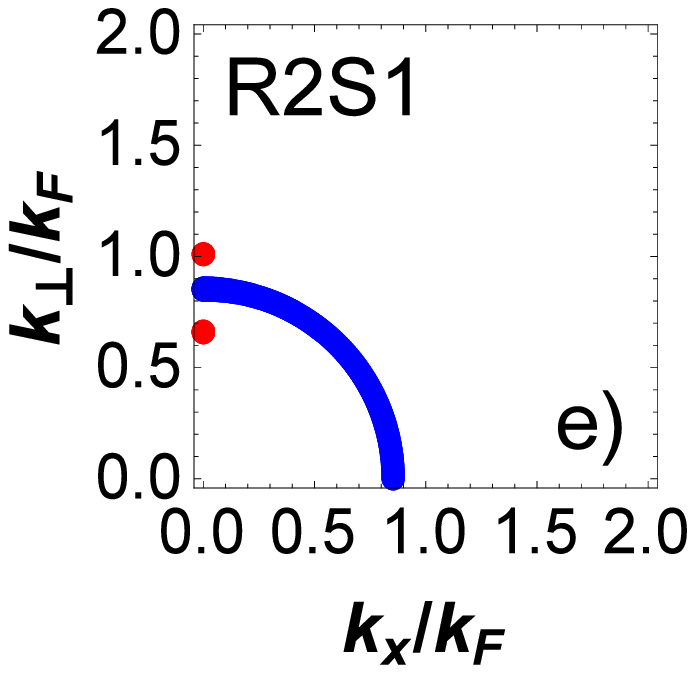,width=0.32 \linewidth}
\caption{ 
\label{fig:two}
(Color Online) 
{\it Loci} of zero quasi-particle energy  
$E_3(k_x,k_{\perp})$ in the $k_xk_\perp$ plane 
at $T = 0.02 E_F$ and $b_z = 0$, 
where $k_{\perp}$ is the radial
component representing $(k_y, k_z)$. 
The phases are: 
a) $N1$ with $\Omega = 1.75E_F$ and $1/(k_F a_s) = -1.00$,
b) $N2$ with $\Omega = E_F$ and $1/(k_F a_s) = -1.00$,
c) $N3$ with $\Omega = 0.40E_F$ and $1/(k_F a_s) = -1.00$,
d) $R1S0$ with $\Omega = 1.75E_F$ and $1/(k_F a_s) = 0.80$,
e) $R1S1$ with $\Omega = E_F$ and $1/(k_F a_s) = 0.20$,
f) $R2S1$ with $\Omega = 0.40E_F$ and $1/(k_F a_s) = -0.30$.
}
\end{figure}

Of the three quasiparticle bands $E_1 ({\bf k})$, $E_2({\bf k})$
and $E_3 ({\bf k})$, only the lowest energy dispersion can have zeros
and therefore be used to classify the emergent superfluid phases based 
on the type of nodal quasiparticles that emerge~\cite{footnote-1}.
Thus, in Fig.~\ref{fig:two},
we plot the momentum space {\it loci} of
$E_3 ({\bf k}) = 0$ versus $(k_x, k_{\perp})$, where $k_{\perp}$
represents a radial vector in the $k_y k_z$ plane,
for zero color shift (detuning $\delta = 0$) and zero quadratic color 
(Zeeman) shift $(b_z = 0)$ describing one the simplest experimental 
situations that can be engineered for 
three internal states of $^{40}$K.
We show only the first quadrant, because the {\it loci} have 
azimuthal symmetry in the  $k_y k_z$ plane and 
reflection symmetry in the $k_x$ direction. 
This means that red dots along the $k_\perp$ axis represent circles in the
$k_y k_z$ plane, and that blue lines in the $k_x k_{\perp}$ represent surfaces
in three-dimensional momentum space $(k_x, k_y, k_z)$.
In the top panels, we describe the normal phases 
$N1, N2, N3$, with one, two or three distinct {\it Fermi} 
surfaces, respectively.
In the bottom panels, we show the nodal structure of three superfluid phases 
that have a boundary with the normal state.
The phase $R1S0$ has one ring and zero
surface of nodes, the phase $R1S1$ has one ring and one surface of nodes, 
and the phases $R2S1$ have two rings and one surface of nodes. 
The phase $R0S1^*$ (not shown in Fig.~\ref{fig:two}) is the limiting case where 
two-rings annihilate in momentum space at the equator of the surface of nodes.  

%
\begin{figure} [t]
\centering
\epsfig{file=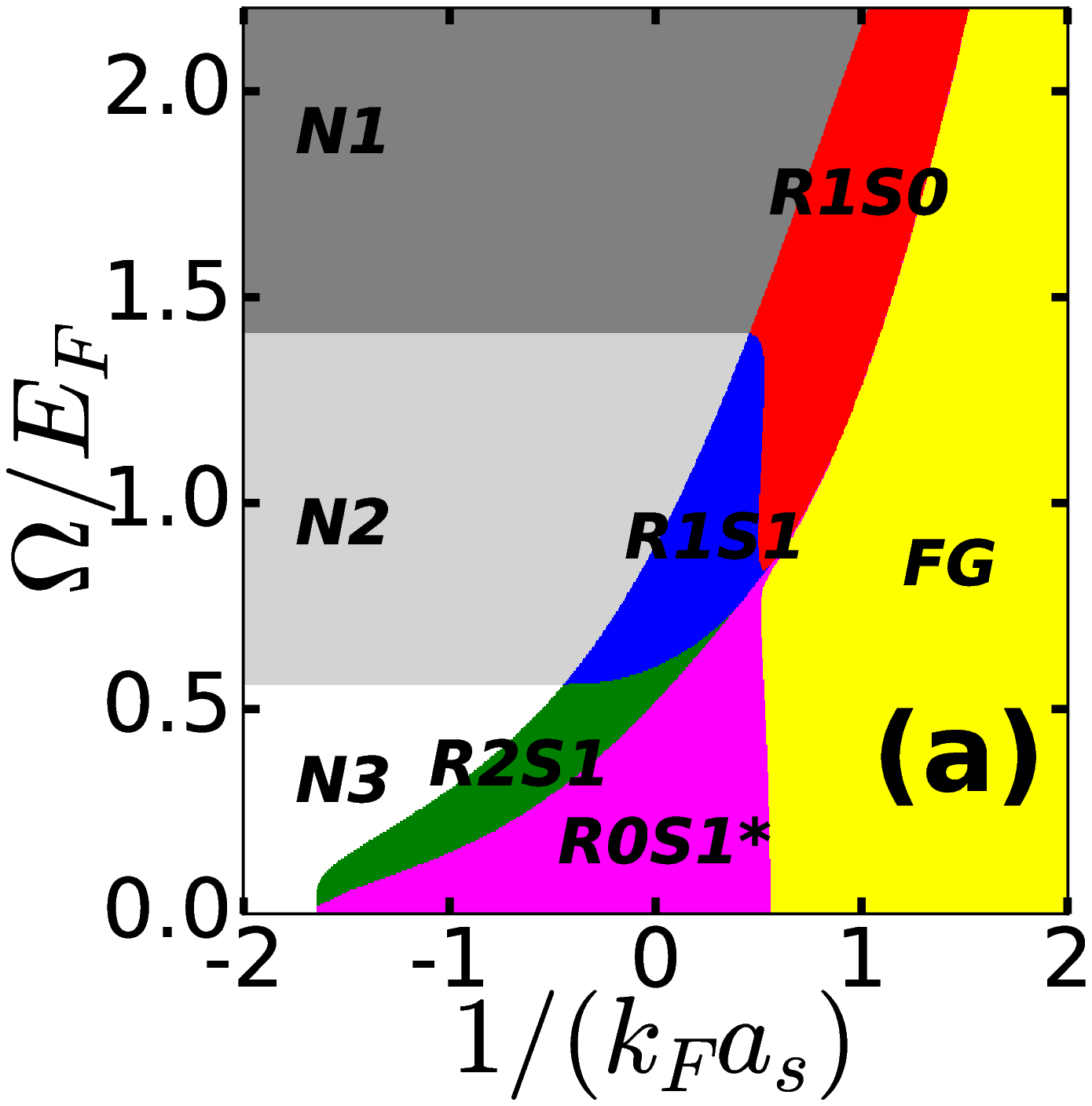,width=0.49 \linewidth}
\epsfig{file=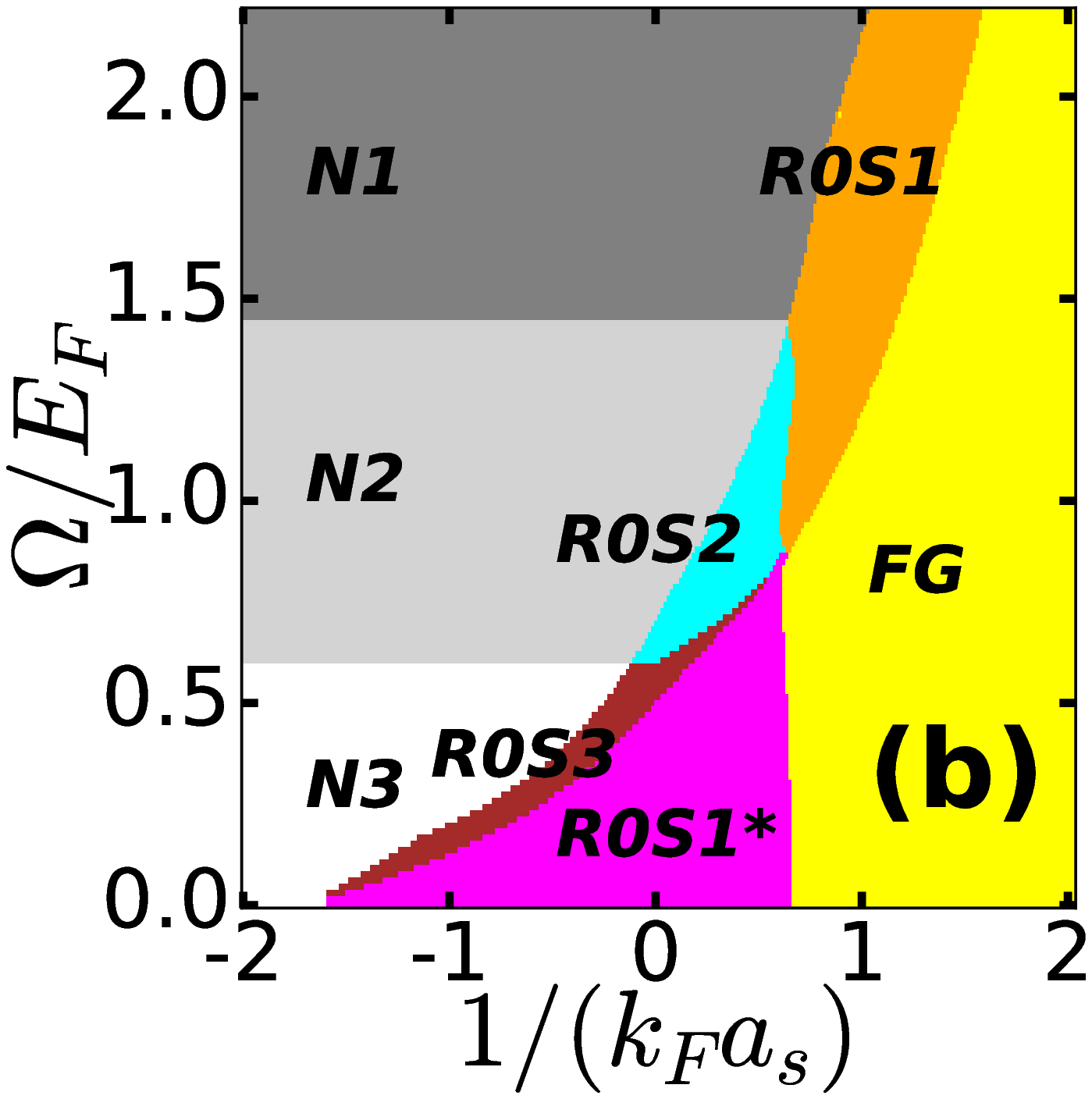,width=0.49 \linewidth}
\caption{ 
\label{fig:three}
(Color Online) 
Phase diagrams of $\Omega/E_F$ versus  $1/(k_F a_s)$  
for $T = 0.02 E_F$ and $b_z = 0$ are shown in 
a) for $k_T = 0.35 k_F$  and in b) for $k_T = 0$. 
The normal phases $N1$, $N2$ and $N3$ 
are indicated by dark-gray, light-gray and white colors, respectively. 
The superfluid phases are color-coded in a) as
$R1S0$ (red), $R1S1$ (blue), 
$R2S1$ (green); 
in b) as 
$R0S1$ (orange) 
$R0S2$ (cyan), 
$R0S3$ (brown);
in a) and b) as 
$R0S1^*$ (magenta), 
$FG$ (yellow).  
}
\end{figure}

In Fig.{~\ref{fig:three}}, we plot phase diagrams 
of Rabi frequency $\Omega/E_F$ versus scattering parameter $1/(k_F a_s)$
for $b_z = 0$ with $k_T = 0.35 k_F$ in Fig.{~\ref{fig:three}}a
and with $k_T = 0$ in Fig.{~\ref{fig:three}}b. 
Normal phases $N1$, $N2$ and $N3$ 
are indicated by dark-gray, light-gray and white colors, respectively.
Gapless superfluid phases are color coded as 
$R1S0$ (red), 
$R1S1$ (blue), 
$R2S1$ (green), 
$R0S1^*$ (magenta)
$R0S1$ (orange), 
$R0S2$ (cyan), 
$R0S3$ (brown), 
and the fully 
gapped $(FG)$ phase (yellow).  
In Fig.~\ref{fig:three}a, where $k_T \ne 0$, 
the phase transitions from superfluid to normal are all continuous. However,
in Fig.~\ref{fig:three}b, where $k_T = 0$, the transition from superfluid 
to normal phases are discontinuous, similar to the case
of two internal states 
with $k_T = 0$~\cite{supplementary-material,seo-2012}.  

The transitions between superfluid phases are topological 
and of the Lifshitz-type~\cite{lifshitz-1960}, where the number of 
simply-connected residual Fermi surfaces change when the phase boundaries
between neighboring superfluid phases are crossed.
In Fig.~\ref{fig:three}(a), there is a quintuple and pentacritical point,
where the phases $R1S0$, $R1S1$, $R2S1$ and $R0S1^*$ 
converge into a fully gapped $FG$ phase. If we start in the $FG$ phase and
circle the quintuple point counterclockwise, a ring emerges 
from ${\bf k} = {\bf 0}$ at the $FG/R1S0$ boundary leading 
to phase $R1S0$, from there a surface of nodes arises from
${\bf k} = {\bf 0}$ at the $R1S0/R1S1$ boundary leading to phase $R1S1$, 
then another ring of nodes appears from ${\bf k} = {\bf 0}$ 
at the $R1S1/R2S1$ boundary. Furthermore, two-rings annihilate at finite 
momentum where a surface of nodes exist at the $R2S1/R0S1^*$ boundary 
leading to phase $R0S1^*$, and finally a full gap emerges at 
the $R0S1^*/FG$ boundary the residual surface of nodes disappear 
through ${\bf k} = {\bf 0}$.
The compressiblity 
$
\kappa = n^2 \left( \partial n / \partial \mu \right)_{T,V},
$
with $n = N/V$, 
becomes 
$
\kappa 
= 
\kappa_c + \kappa_p 
\vert \tilde \mu - \tilde \mu_c\vert^{1/2}
$
when the critical point is approached through phases with 
dominant nodal surfaces, or 
becomes
$
\kappa 
= 
\kappa_c 
+ 
\kappa_{R1S0} \vert \tilde \mu - \tilde \mu_c \vert
$
when the critical point is approached through the single ring 
phase $R1S0$. Here, $\tilde \mu = \mu/E_F$, $\mu_c$ 
is the critical chemical potential, and $\kappa_p$ is a phase
dependent coefficient. In all cases, the derivative
$
\left( \partial \kappa / \partial \mu\right)_{T,V}
$ 
is discontinuous 
at the critical point.
Similarly, at the pentacritical point 
in Fig.~\ref{fig:three}b, the compressibility 
$\kappa = \kappa_c + \kappa_p \vert \tilde \mu - \tilde \mu_c \vert^{1/2}$,
when approached from gapless phases with surface nodes. 

To understand the emergence of  momentum dependence in the order parameter,
we write
${\bf H}_{\rm MF} ({\bf k})$ as 
\begin{eqnarray}
\label{eqn:mean-field-hamiltonian-matrix-helicity}
\widetilde{\bf H}_{\rm MF}({\bf k})
=
\left(
\begin{array}{cc}
\widetilde {\bf H}_D ({\bf k})	&	\widetilde {\bf \Lambda}
\\
\widetilde{\bf \Lambda}^{\dagger} &	-\widetilde {\bf H}_D^* (-{\bf k})
\end{array}
\right)
\end{eqnarray}
in the mixed-color basis $\alpha = \{\Uparrow, 0, \Downarrow \}$,
where 
$
{\bf \Phi}^{\dagger}_{\bf k} 
= 
\left[
\phi^{\dagger}_{\Uparrow} ({\bf k}),
\phi^{\dagger}_{0} ({\bf k}),
\phi^{\dagger}_{\Downarrow} ({\bf k})
\right]
$ 
is related to the $\{R,G, B\}$ 
basis ${\bf \Psi}^\dagger_{\bf k}$ via a unitary 
transformation
$
{\bf \Phi}^\dagger_{\bf k} 
= 
{\bf \Psi}^\dagger_{\bf k}
{\bf U}^\dagger ({\bf k}).
$
The matrix elements of the $3 \times 3$ blocks are 
$
\widetilde {\bf H}_{D,\alpha \beta} ({\bf k})
=
\widetilde{\mathcal E}_{\alpha} ({\bf k})
\delta_{\alpha \beta},
$ 
with 
$
\widetilde {\mathcal E}_{\alpha} ({\bf k})
=
{\mathcal E}_{\alpha} ({\bf k})
-
\mu
$
and 
$
\widetilde {\bf \Lambda}_{\alpha \beta}
=
\Delta_{\alpha \beta} ({\bf k}).
$	
The matrix 
$
\widetilde {\bf \Lambda}
$
describing the order parameter tensor
$
\Delta_{\alpha \beta} ({\bf k})
$
is momentum {\it dependent} in contrast
to the original matrix
$
{\bf \Lambda}, 
$
which is {\it independent} of momentum. The order parameter tensor 
becomes
\begin{equation}
\label{eqn:order-parameter-tensor-mixed-color-basis}
\Delta_{\alpha \beta} ({\bf k})
=
\Delta 
\left[
u_{\alpha R} ({\bf k}) u_{\beta B} ({\bf - k})
-
u_{\alpha B} ({\bf k}) u_{\beta R} ({\bf - k})
\right],
\end{equation}
where $u_{\alpha c} ({\bf k})$ are matrix elements
of ${\bf U}({\bf k})$ 
and represent the $R$ and $B$ 
components of the eigenvector amplitudes
${\bf u}_{\alpha} ({\bf k}) 
= 
\left[
u_{\alpha R} ({\bf k}), u_{\alpha G} ({\bf k}), u_{\alpha B} ({\bf k})
\right].
$
The property 
$
\Delta_{\alpha \beta} ({\bf k}) 
= 
- \Delta_{\beta \alpha} ({\bf - k})
$
guarantees that the diagonal elements
$\Delta_{\alpha \alpha} ({\bf k})$ 
have odd parity due to the Pauli principle. 
Also, $\Delta_{\alpha \beta} ({\bf k})$ has nine components and 
can be written in the basis of total pseudo-spin $S$ and total 
pseudo-spin projection $m_s$ with singlet $(S = 0)$, triplet $(S = 1)$ 
and quintuplet $(S = 2)$ sectors. This is achieved by 
writing ${\widetilde \Delta}_{S m_s} ({\bf k}) = M_{\alpha \beta}^{S m_s}
\Delta_{\alpha \beta} ({\bf k})$, where $M_{\alpha \beta}^{S m_s}$ is a 
tensor of generalized 
Clebsch-Gordon coefficients~\cite{supplementary-material}.

%
\begin{figure} [t]
\centering 
\epsfig{file=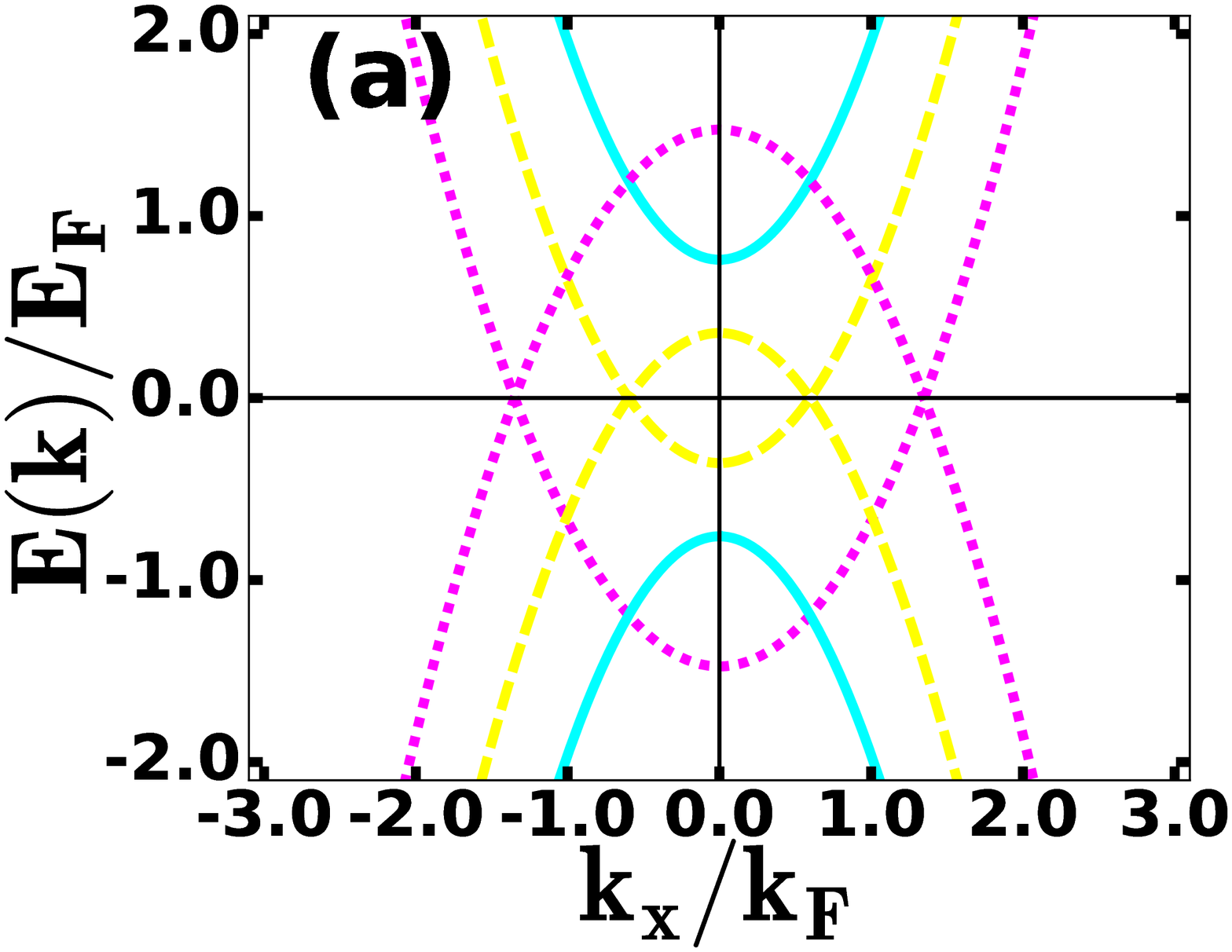,width=0.49 \linewidth}
\epsfig{file=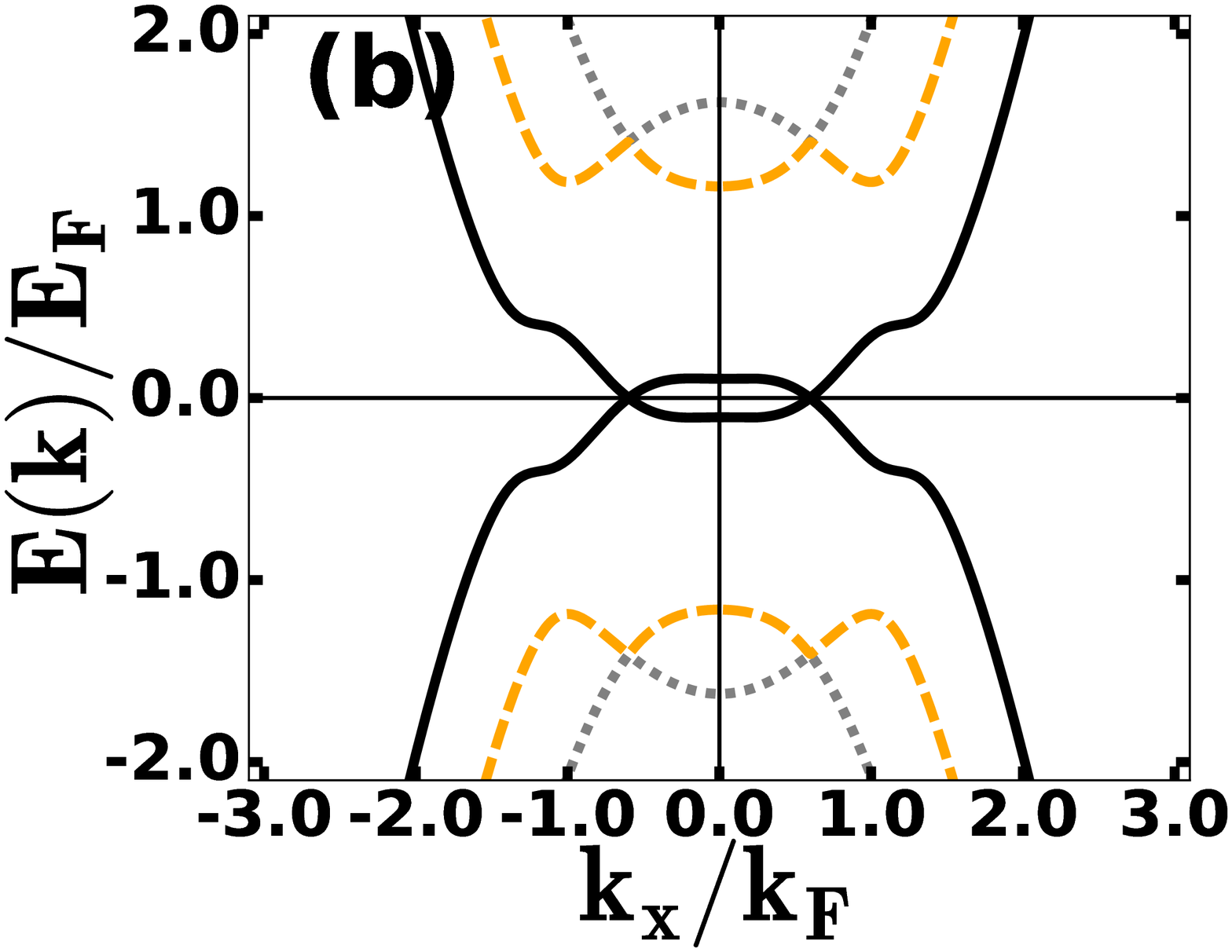,width=0.49 \linewidth}
\\
\epsfig{file=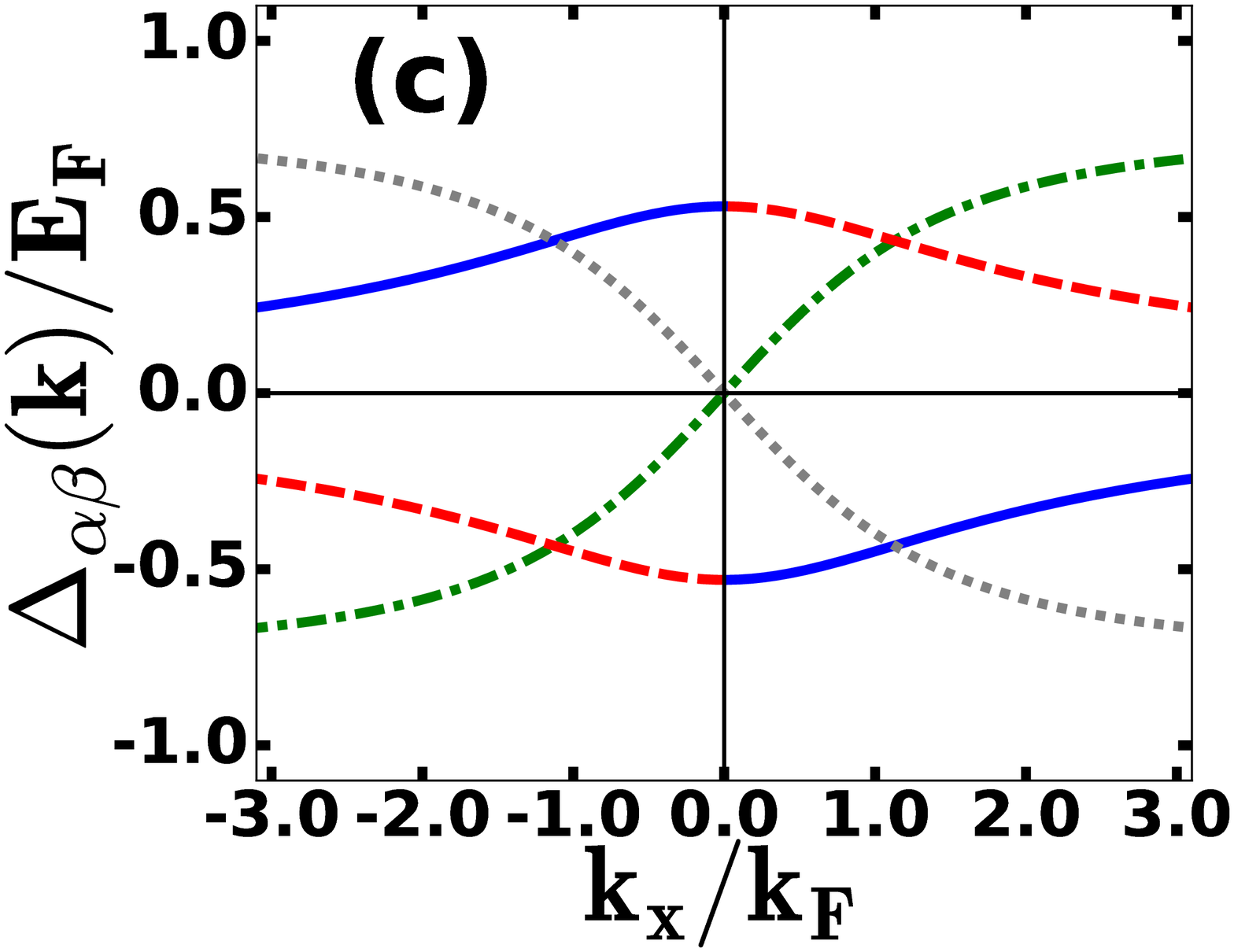,width=0.49 \linewidth}
\epsfig{file=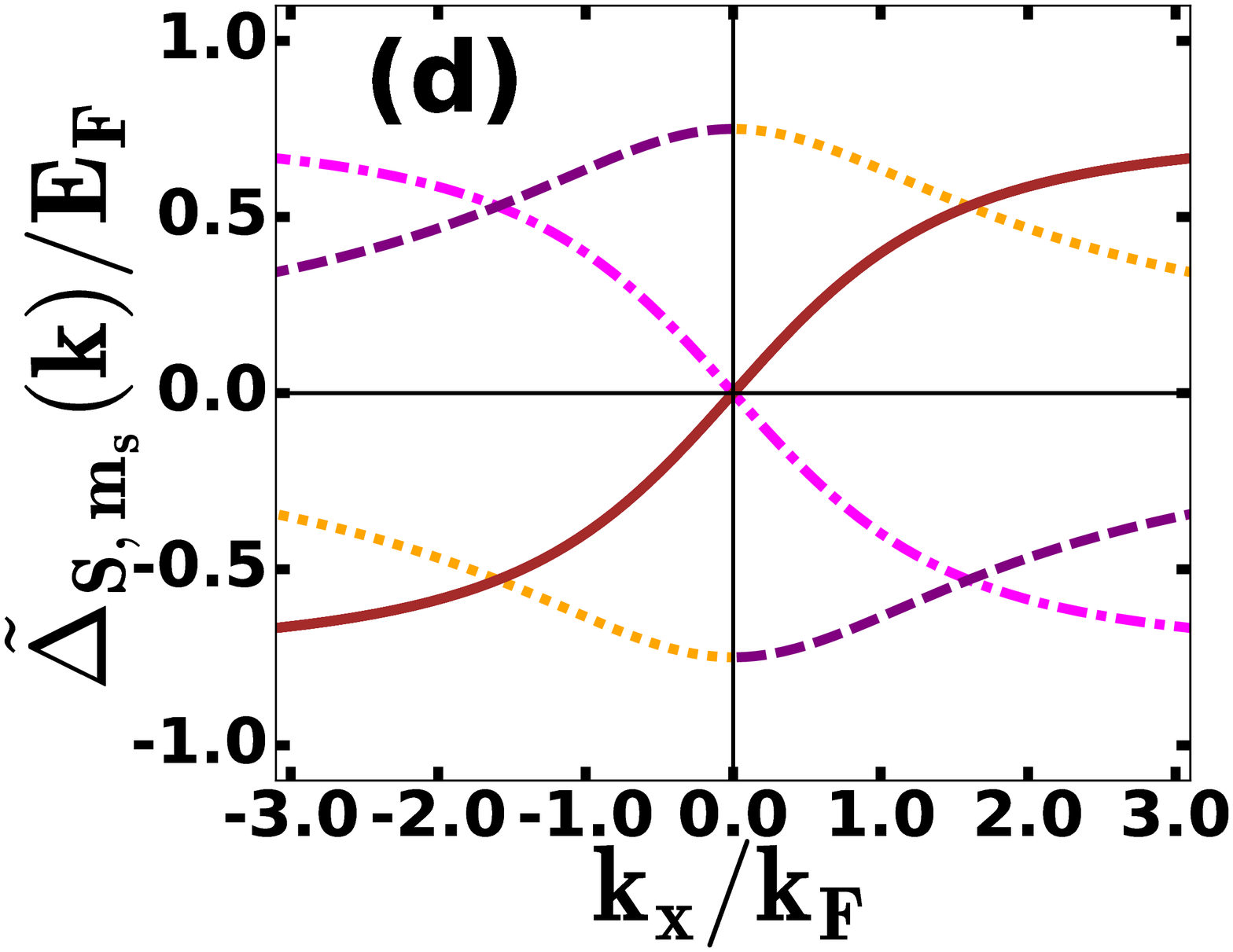,width=0.49 \linewidth}
\caption{ 
\label{fig:four}
(Color online) All panels have $k_T/k_F = 0.35$ and $T/E_F = 0.02$.
a) Mixed-color particle and hole energies 
$\pm \xi_{\Uparrow} ({\bf k})$ (dotted magenta), 
$\pm\xi_0 ({\bf k})$, (dashed yellow) 
$\pm \xi_{\Downarrow} ({\bf k})$ (solid cyan)
versus momentum $k_x$ for the normal phase $N2$; 
where $\Delta/E_F  = 0$, $\mu/E_F = 0.73$,
$\Omega/E_F = 0.79$ and $1/(k_F a_s) = -1.5$.
Panels b), c) and d) refer to superfluid phase
$R1S1$ with parameters $\Delta/E_F = 0.75$, $\mu/E_F = 0.48$, 
$\Omega/E_F = 0.79$, and  $1/(k_F a_s) = 0.23$.
b) Quasiparticle $(E ({\bf k}) > 0)$ and quasihole $(E({\bf k}) < 0)$ 
energies  versus $k_x$, 
$\pm E_1 ({\bf k})$ (dotted gray), 
$\pm E_2 ({\bf k})$ (dashed orange)
$\pm E_3 ({\bf k})$ (solid black). 
c) Order parameter matrix elements in mixed-color basis 
$\Delta_{\Uparrow\Uparrow} ({\bf k})$ (dot-dashed green), 
$\Delta_{\Uparrow 0} ({\bf k})$ (solid blue),
$\Delta_{\Downarrow\Downarrow} ({\bf k})$ (dotted gray), and
$\Delta_{\Downarrow 0} ({\bf k})$ (red dashed).
d) Order parameter matrix elements in quintuplet sector $(S = 2)$
$\Delta_{22} ({\bf k})$ (solid brown),
$\Delta_{21} ({\bf k})$ (dot-dashed magenta),
$\Delta_{20} ({\bf k}) = 0$,
$\Delta_{2\bar1} ({\bf k})$ (dotted orange),
$\Delta_{2\bar2} ({\bf k})$ (dashed purple).
}
\end{figure}

For instance, in Fig.~\ref{fig:three}a ($k_T \ne 0$), 
when $\Omega \ne 0$, all
superfluid phases have three vanishing order parameter components 
$
\Delta_{00} ({\bf k}) 
= \Delta_{\Uparrow \Downarrow} ({\bf k})
= \Delta_{\Downarrow \Uparrow} ({\bf k}) 
= 0
,
$ 
while the six remaining 
non-vanishing components can be obtained from 
$\Delta_{\Uparrow 0} ({\bf k})$ and $\Delta_{\Uparrow \Uparrow} ({\bf k})$
via the symmetry relations:
$\Delta_{0 \Uparrow} ({\bf k}) = \Delta_{\Uparrow 0} ({\bf k})$;
$\Delta_{\Downarrow 0} ({\bf k}) = - \Delta_{\Uparrow 0} ({\bf k})$;
$\Delta_{\Downarrow \Downarrow} ({\bf k}) = 
-\Delta_{\Uparrow \Uparrow} ({\bf k})$;
$\Delta_{0 \Downarrow} ({\bf k}) = \Delta_{\Downarrow 0} ({\bf k})$.
For all superfluid phases of Fig.~\ref{fig:three}a with 
$\Omega \ne 0$, only the quintuplet sector $(S = 2)$ has 
non-vanishing components, thus leading to very unconventional color 
pairing, beyond the singlet and triplet channels of 
spin-$1/2$ fermions in condensed matter physics. 
When $\Omega = 0$, as discussed earlier, 
the system trivializes since there is no color-mixing and no 
momentum dependence in the order parameter. 

In Fig.~\ref{fig:four}a, we show the mixed-color particle $(+)$ and 
hole $(-)$ energies $\pm \xi_{\Uparrow} ({\bf k}), \pm\xi_0 ({\bf k}), 
\pm \xi_{\Downarrow} ({\bf k})$ versus momentum $k_x$ 
for the normal phase $N2$, where $\Delta  = 0$, 
with $\Omega = 0.79$ and $1/(k_F a_s) = -1.5$.
In Fig.~\ref{fig:four}b, we show the quasi-particle (positive) energies
$E_{1} ({\bf k})$ (dotted gray),
$E_{2} ({\bf k})$ (dashed orange)
$E_{3} ({\bf k})$ (black solid)
and quasi-hole (negative) energies
$E_4 ({\bf k}) = -E_3 (-{\bf k})$ (black solid)
$E_5 ({\bf k}) = -E_2 (-{\bf k})$ (orange dashed)
$E_6 ({\bf k}) = -E_1 (-{\bf k})$ (gray dotted)
energies 
for the superfluid phase $R1S1$, 
with $\Omega/E_F = 0.79$ and $1/(k_F a_s) = 0.23$. 
In Fig.~\ref{fig:four}c, we show the order parameter components 
$\Delta_{\Uparrow\Uparrow} ({\bf k})$ (dot-dashed green), 
$\Delta_{\Uparrow 0} ({\bf k})$ (solid blue),
$\Delta_{\Downarrow\Downarrow} ({\bf k})$ (dotted gray), and
$\Delta_{\Downarrow 0} ({\bf k})$ (dashed red). 
The gaps
in the excitation spectrum of Fig.~\ref{fig:four}b occur at momentum 
locations where $\Delta_{\alpha \beta} ({\bf k})$ lift the degeneracies of 
the particle and hole energies shown in Fig.~\ref{fig:four}a.
In Fig.~\ref{fig:four}d, we show the components
$\Delta_{22} ({\bf k})$ (solid brown),
$\Delta_{21} ({\bf k})$ (dot-dashed magenta),
$\Delta_{20} ({\bf k}) = 0$,
$\Delta_{2\bar1} ({\bf k})$ (dotted orange),
$\Delta_{2\bar2} ({\bf k})$ (dashed purple)
in quintuplet sector $(S = 2)$.

In conclusion, we proposed the existence of unconventional color 
superfluids with quintuplet pairing in the presence of color-orbit 
coupling for ultra-cold fermions. When s-wave interactions 
and color-flip fields are changed, we found that the resulting phase diagram 
is very rich, containing a quintuple and pentacritical point, where 
the compressibility is non-analytic and four gapless superfluid phases 
converge into a fully gapped one. 

\acknowledgments{One of us (C. A. R. SdM) acknowledges support from
the Joint Quantum Institute via a Sabbatical Visit, 
the Galileo Institute for Theoretical Physics via a Simons Fellowship, 
and the International Institute of Physics via it's Visitors 
Program.}

\pagebreak

\widetext
\begin{center}
\vskip 0.3cm
\textbf{ \large Supplementary Material}
\vskip 0.3cm 
\textbf{ \large
Unconventional color superfluidity in ultra-cold fermions: \\
Quintuplet pairing, quintuple point and pentacriticality \\
}
\vskip 0.3cm
{
Doga Murat Kurkcuoglu and C. A. R. S{\'a} de Melo \\
{\it School of Physics, Georgia Institute of Technology, Atlanta, 
Georgia 30332, USA}
}
\end{center}

\vskip 0.3cm

Heating effects within Raman schemes have so far precluded 
studies of ultra-cold fermions with spin-orbit coupling at temperatures
$T < 0.3 E_F$, where $E_F$ is the system's Fermi energy. 
However, new techniques using radio-frequency chips are under
development to produce spin-orbit and color-orbit couplings without 
the use of Raman lasers and thus without heating the system. 
For ultra-cold fermions with three colors (internal
states) $\{ R, G, B \}$, a generic situation can be created 
via radio-frequency chips or Raman lasers leading to the 
independent particle Hamiltonian matrix
\begin{eqnarray}
\label{eqn:independent-particle-hamiltonian-matrix}
{\bf H}_0({\bf k})
= 
\left(
\begin{array}{ccc}
\varepsilon_R({\bf k}) 	& \Omega_{RG}	         &   \Omega_{RB}	  \\
\Omega_{RG}^* 		& \varepsilon_G({\bf k}) &   \Omega_{GB}	  \\
\Omega_{RB}^* 		& \Omega_{GB}^*	         & \varepsilon_B({\bf k}) \\
\end{array} 
\right),
\end{eqnarray}
where 
$
\varepsilon_c({\bf k}) = ({\bf k}-{\bf k}_c)^2/(2m) + \eta_c
$
represents the energy of internal color state 
$
c = 
\{
R, G, B
\}
$
after net momentum transfer
${\bf k}_c$, and $\eta_c$ is 
a reference energy of the atom at internal state $c$. 
The matrix elements $\Omega_{c c^{\prime}}$ represent Rabi frequencies
between atomic color states $c$ and $c^{\prime}$. 

In the main text, we investigate a simpler experimental situation by setting the Rabi 
frequency $\Omega_{RB} = 0$ 
in Eq.~(\ref{eqn:independent-particle-hamiltonian-matrix}),
indicating that there is no coupling between states
$R$ and $B$. In addition, we consider that the Rabi frequencies associated
with the transitions from states $R$ to $G$ and from $R$ to $B$ to be real and 
equal, that is, 
$\Omega_{RG} = \Omega_{RG}^* = \Omega_{GB} = \Omega_{GB}^* = \Omega$.
Furthermore, we assume that momentum transfers occur only in states 
$R$ and $B$, such that 
${\bf k}_R = k_T {\hat {\bf x}}$,
${\bf k}_G = 0$,
and ${\bf k}_B = -k_T {\hat {\bf x}}$,
where $k_T$ is the magnitude of the momentum transferred 
to the atom by photons or radio-frequency fields.
Lastly, we can define an energy reference via the sum 
$
\sum_c \eta_c = \eta,
$
leading to internal energies 
$
\eta_R = -\delta,
$
$
\eta_G = \eta
$ 
and
$
\eta_B = +\delta,
$
where $\delta$ represents the detuning. Under these considerations
the Hamiltonian matrix ${\bf H}_0 ({\bf k})$ shown above acquires the form
described in Eq.~(\ref{eqn:color-hamiltonian}) of the main text, when 
written in terms of the spin-one angular momentum matrices ${\bf J}_{\ell}$ 
with $\ell = \{ x, y, z \}$.  

In second quantization, the Hamiltonian matrix ${\bf H}_0 ({\bf k})$ becomes 
the chip-atom (Raman-atom) Hamitonian
\begin{equation}
\label{eqn:chip-atom-hamiltonian-supplementary-material}
{H}_{{\rm CA}}
= 
\sum_{\bf k}
{\bf \Psi}^{\dagger}_{\bf k}
{\bf H}_0({\bf k})
{\bf \Psi}_{\bf k},
\end{equation}
where the spinor creation operator is 
$
{\bf \Psi}^{\dagger}_{\bf k} 
= 
\left[
\psi_R^{\dagger}({\bf k}),
\psi_G^{\dagger}({\bf k}),
\psi_B^{\dagger}({\bf k})
\right]
$,
with
$
\psi_c^{\dagger}({\bf k})
$
creating a fermions labelled by momentum 
${\bf k}$ and color $c = \{ R, G, B \}$. 
The Hamiltonian $H_{\rm CA}$
can be diagonalized via the rotation 
$
{\bf \Phi} ({\bf k}) 
= 
{\bf U}({\bf k}) 
{\bf \Psi} ({\bf k}),
$
which connects the three-component spinor ${\bf \Psi} ({\bf k})$ 
in the original color basis to the three-component spinor 
${\bf \Phi} ({\bf k})$ 
representing the mixed color basis, that is, the basis 
of eigenstates. The matrix ${\bf U}({\bf k})$ 
is unitary and satisfies the relation 
${\bf U}^\dagger ({\bf k}) {\bf U} ({\bf k}) = {\bf 1}$. 
The diagonalized Hamiltonian matrix is
\begin{equation}
\label{eqn:diagonal-hamiltonian-supplementary-material}
{\bf H}_D({\bf k}) 
= 
{\bf U}({\bf k})
{\bf H}_{0}({\bf k}) 
{\bf U}^{\dagger} ({\bf k}),
\end{equation}
with matrix elements 
$
{\bf H}_{D, \alpha \beta} ({\bf k}) 
=
{\mathcal E}_{\alpha} ({\bf k}) \delta_{\alpha \beta},
$ 
where $\mathcal{E}_{\alpha} ({\bf k})$ are the eigenvalues of 
${\bf H}_{0} ({\bf k})$ in Eq.~(\ref{eqn:color-hamiltonian}) of the main text.
The three-component spinor in the mixed-color eigenbasis is  
$
{\bf \Phi}^{\dagger} ({\bf k}) 
= 
\left[
\phi^{\dagger}_{\Uparrow} ({\bf k}),
\phi^{\dagger}_{0} ({\bf k}),
\phi^{\dagger}_{\Downarrow} ({\bf k})
\right],
$ 
where $\phi^{\dagger}_{\alpha} ({\bf k})$ is the creation 
operator of a fermion with eigenenergy $\mathcal{E}_{\alpha} ({\bf k})$
and mixed-color label $\alpha$.
The unitary matrix
\begin{eqnarray}
\label{eqn:unitary-matrix}
{\bf U}({\bf k}) 
= 
\left(
\begin{array}{c c c}
u_{\Uparrow R} ({\bf k}) & u_{\Uparrow G} ({\bf k}) & u_{\Uparrow B} ({\bf k}) \\
u_{0 R} ({\bf k}) & u_{0 G} ({\bf k}) & u_{0 B} ({\bf k}) \\
u_{\Downarrow R} ({\bf k}) & u_{\Downarrow G} ({\bf k}) & u_{\Downarrow B} ({\bf k})  
\end{array}
\right)
\end{eqnarray}
has rows that satisfy the normalization condition
$
\sum_{c}
\vert u_{\alpha c} ({\bf k}) \vert^2 
= 
1,
$
where $\alpha = \{\Uparrow, 0, \Downarrow \}$.

When $b_z = 0$, the Hamiltonian matrix
${\bf H}_0 ({\bf k})$ in Eq.~(\ref{eqn:color-hamiltonian}) of the main text reduces 
to that of spin-one fermions under a momentum-dependent magnetic field. 
In this case, the eigenvalues of ${\bf H}_0 ({\bf k})$ are 
\begin{equation}
\label{eqn:eigenvalues-chip-atom-hamiltonian-supplementary-material}
{\mathcal E}_\alpha ({\bf k}) 
= 
\varepsilon ({\bf k}) 
- 
 m_\alpha \vert h_{\rm eff} ({\bf k}) \vert,
\end{equation}
with $m_\alpha = \{+1, 0,-1\}$.
Here, the reference kinetic energy
$
\varepsilon ({\bf k}) = {\bf k}^2/(2m) + \eta
$ 
is the same for all colors with $\eta = k_T^2/2m$, and the effective
momentum-dependent magnetic field amplitude is 
\begin{equation}
\label{eqn:effective-magnetic-field}
\vert h_{\rm eff} ({\bf k}) \vert 
=
\sqrt{
\vert h_x ({\bf k}) \vert^2 
+ 
\vert h_z({\bf k}) \vert^2,
}
\end{equation}
where
$
h_x({\bf k}) = -\sqrt{2} \Omega
$ 
is the color-flip Rabi field, and 
$
h_z({\bf k}) = 2k_T k_x /(2m) + \delta
$
is a momentum dependent Zeeman field along the $z$-axis.
As mentioned above, the label $\alpha$ 
describes mixed color states induced by color-orbit coupling via the
color-dependent momentum transfer $k_T$ and by color-flip fields via
the Rabi frequency $\Omega$.

Adding attractive contact interactions 
$-g_{cc^{\prime}} \delta ({\bf r} - {\bf r}^{\prime})$ of strength 
$g_{c c^{\prime}} > 0$ between internal states $c \ne c^{\prime}$, 
and focusing on uniform superfluid phases
${\bf Q} = {\bf 0}$ with order parameter tensor 
$
\Delta_{cc^{\prime}}
= 
-g_{cc^{\prime}}
\langle 
b_{cc^{\prime}}
(
{\bf 0}
)
\rangle/V,
$
leads to the mean-field Hamiltonian
\begin{equation}
\label{eqn:mean-field-Hamiltonian-supplementary-material}
H_{\rm MF} 
= 
\frac{1}{2}
\sum_{\bf k}
{\bf \Psi}^{\dagger}_{N} ({\bf k})
{\bf H}_{\rm MF}({\bf k})
{\bf \Psi}_{N} ({\bf k})
+
V \sum_{c\neq c^{\prime}}
\frac{|\Delta_{cc^{\prime}}|^2}{g_{cc^{\prime}}}
+
{\mathcal C} (\mu) 
\end{equation}
described in Eq.~(\ref{eqn:mean-field-Hamiltonian}) of the main text.
Here, 
$
{\bf \Psi}^{\dagger}_{N} ({\bf k})
=
\left[
\Psi_R^{\dagger}({\bf k}),
\Psi_G^{\dagger}({\bf k}),
\Psi_B^{\dagger}({\bf k}),
\Psi_R ({-\bf k}),
\Psi_G ({-\bf k}),
\Psi_B ({-\bf k})
\right]
$
is a six-component Nambu spinor, and $b_{cc^\prime} ({\bf 0})$ is the 
pairing operator with zero center of mass momentum. As indicated in the 
main text, we particularized our discussion to the case where s-wave 
interactions exist only between states $\vert R\rangle$
and $\vert B\rangle$, such that the $\vert G \rangle$ state does not 
interact with the other two.  In this situation, the order parameter 
tensor $\Delta_{c c^\prime}$ is represented by a single 
scalar $\Delta_{RB} = \Delta$.

The Hamiltonian matrix ${\bf H}_{\rm MF} ({\bf k})$ in 
Eq.~(\ref{eqn:mean-field-hamiltonian-matrix}) of the main text 
has six eigenvalues, which we order as  
$
E_1({\bf k})
>
E_2({\bf k})
>
E_3({\bf k})
>
E_4({\bf k})
>
E_5({\bf k})
>
E_6({\bf k}).
$
These eigenvalues exhibit quasiparticle/quasihole symmetry
in momentum space for any value of detuning $\delta$ and Rabi 
frequency $\Omega$, which means 
$
E_6({\bf k}) = -E_1(-{\bf k}),
$
$
E_5({\bf k}) = -E_2(-{\bf k})
$
and
$
E_4({\bf k}) = -E_3(-{\bf k}).
$
However, each eigenergy $E_j ({\bf k})$ has well defined parity only
when $\delta = 0$, in which case $E_j ({\bf k}) = E_j ({\bf - k})$ 
has even parity. In the limit of zero color-flip field $\Omega \to 0$, 
with $k_T = 0$ or $k_T \ne 0$, the excitation spectrum can be 
obtained analytically as the bands describing states 
$\{ R, G, B \}$ do not mix, in which case, state 
$\vert G \rangle$ is completely inert to pairing. When $\Omega = 0$, 
a spin-gauge symmetry relates trivially the states with 
$k_T = 0$ and $k_T \ne 0$, that is, $k_T$ can be gauged away.  
This leads to quasiparticle energies $E_1 ({\bf k}) = E_2 ({\bf k}) = 
\sqrt{ \xi_{\bf k}^2  + \vert \Delta \vert^2 }$, where 
$E_3 ({\bf k})$ reduces to $\vert \xi_{\bf k} \vert$, with  
$\xi_{\bf k} = \epsilon_{\bf k} - \mu $ 
being the independent particle energy.
Thus, s-wave pairing occurs only between $R$ and $B$ states,
while there is no pairing involving the $G$ state, as expected. 
The only physical role played by state $G$ is to contribute to the total 
density of fermions, and thus to affect the chemical potential $\mu$.
Furthermore, when $\Omega \to 0$, the number of particles in each band 
is conserved separately and color-orbit coupling can be gauged away, 
leading to an inert band $G$ and to standard BCS-BEC crossover phenomena in 
the superfluid phase for bands $R$ and $B$.

To analyze the excitation spectrum $E_j ({\bf k})$, we need to determine
self-consistently the values of the order parameter amplitude 
$\Delta_{RB} = \Delta$ and the chemical potential $\mu$. 
For this purpose, we write the thermodynamic potential
$
\mathcal{Q}
= 
-T
\ln
\mathcal{Z},
$
where 
$
\mathcal{Z} 
= 
\int
\Pi_{s}D
\left[
\psi^{\dagger}_c ({\bf k})
,
\psi_c ({\bf k})
\right]
\exp
\left[
-S
\right]
$
is the grand-canonical partition function written in terms of the
action $S$. At the mean-field level the action is
\begin{equation}
\label{eqn:mean-field-action}
T S_{\rm MF} 
= 
-
\frac{1}{2}
\sum_{n,{\bf k} }
{\bf \Psi}_{N}^{\dagger} ({\bf k})
{\bf G}^{-1} 
{\bf \Psi}_{N} ({\bf k})
+
V \frac{\vert \Delta_{RB} \vert^2}{g_{RB}}
+
{\mathcal C} (\mu), 
\end{equation}
where 
$
{\bf G}^{-1} (i\omega_n, {\bf k})
=
\left[
i\omega_n {\bf 1}
-
{\bf H}_{\rm MF}({\bf k})
\right]
$
is the inverse of the resolvent (Green) matrix 
${\bf G} (i\omega_n, {\bf k})$. Here, 
$
\omega_n = (2n+1)\pi T
$
is the fermionic Matsubara frequency and 
$T$ is the temperature. A Gaussian integration over the 
fermionic fields 
$\{ {\bf \Psi}_N^\dagger ({\bf k}),  {\bf \Psi}_N ({\bf k}) \}$
leads to the thermodynamic potential ${\cal Q}_{\rm MF}$ shown in 
Eq.~(\ref{eqn:thermodynamic-potential}) of the main text, 
from which the self-consistent order parameter and number equations can 
be obtained as Eqs.~(\ref{eqn:order-parameter-equation}) 
and~(\ref{eqn:number-equation}) in the main text.

The transitions between normal and superfluid phases
are continuous for $\Omega \ne 0$ and $k_T \ne 0$ 
and are discontinuous for $\Omega \ne 0$ and $k_T = 0$ in the
cases of the phase diagrams shown in Figs.~\ref{fig:three}a 
and~\ref{fig:three}b of the main text, respectively. 
For three color states with single channel interaction $g_{RB} = g$ 
and $k_T = 0$, the mixing of color states induced by the color-flip 
field (Rabi frequency) $\Omega$ does not smooth out the transition into 
a continous one. This occurs because $\Omega$
mix states $\vert R\rangle$, $\vert G\rangle$ and $\vert B\rangle$, 
such that the only non-vanishing components of the order parameter tensor 
$\Delta_{\alpha \beta} ({\bf k})$ 
in the mixed color basis, 
Eq.~(\ref{eqn:order-parameter-tensor-mixed-color-basis}) of the main text,
are momentum independent and equal to 
$
\Delta_{{\Uparrow} 0} ({\bf k})
= 
\Delta_{0 \Downarrow} ({\bf k})
= 
-\Delta_{0 \Uparrow} ({\bf k}) 
= 
-\Delta_{\Downarrow 0} ({\bf k})  
= 
\Delta_{RB}~{\rm sgn}[\Omega]/\sqrt{2}.
$ 
Thus, there is an energy cost for pairing the $\vert R\rangle$ 
and $\vert B \rangle$ components of mixed color states
$\vert 0\rangle$ and $\vert \Uparrow \rangle$ or $\vert 0 \rangle$ and 
$\vert \Downarrow \rangle$ with zero center 
of mass momentum, similar to what happens for s-wave pairing for spin-1/2 
systems in a Zeeman field. This conclusion is supported by an 
analysis of the thermodynamic 
potential ${\cal Q}$. The thermodynamic potential 
difference between the superfluid and normal states is
$
{\cal \delta Q} = {\cal Q} - {\cal Q}_N = a \vert \Delta \vert^2 
+ b \vert \Delta \vert^4 + c \vert \Delta \vert^6    
$
near the normal-superfluid phase boundary, 
where the parameters $b$ and $c$ are always positive over the entire phase 
space while $a$ changes sign when the phase boundary is crossed 
for $k_T \ne 0$. However, for $k_T = 0$, $b$ may change sign, but $c$ remains
positive, leading to a discontinuous transition. Such discontinuity, occurs
at the Clogston limit obtained by balancing the {\it magnetic} energy 
$
h_x \chi_{xx} h_x/2
$ 
and the condensation energy 
$\gamma \vert \Delta \vert^2$, leading to the phase
boundary 
$
\Omega = \vert \Delta \vert \gamma/\chi_{xx},
$ 
where $\vert \Delta \vert$ jumps 
discontinuously to zero.

The order parameter $\Delta_{\alpha \beta} ({\bf k})$, defined in 
Eq.~(\ref{eqn:order-parameter-tensor-mixed-color-basis}) of the main text,
has nine components and can be written in the basis of total pseudo-spin $S$ 
and total pseudo-spin projection $m_s$ with singlet $(S = 0)$, 
triplet $(S = 1)$  and quintuplet $(S = 2)$ sectors. This is achieved by 
writing ${\widetilde \Delta}_{S m_s} ({\bf k}) = M_{\alpha \beta}^{S m_s}
\Delta_{\alpha \beta} ({\bf k})$, where $M_{\alpha \beta}^{S m_s}$ is a 
tensor of generalized Clebsch-Gordon coefficients. The explicit forms of
these matrix elements are 
\begin{equation}
{\widetilde \Delta}_{00} ({\bf k}) 
= 
\frac{1}{\sqrt{3}} \Delta_{\Uparrow\Downarrow} ({\bf k})
-
\frac{1}{\sqrt{3}} \Delta_{00} ({\bf k})
+
\frac{1}{\sqrt{3}} \Delta_{\Downarrow\Uparrow} ({\bf k}),
\end{equation}
for the singlet sector with $S = 0$, with $m_s = 0$;
\begin{eqnarray}
{\widetilde \Delta}_{11} ({\bf k}) 
& = &
\frac{1}{\sqrt{2}} \Delta_{\Uparrow 0} ({\bf k})
-
\frac{1}{\sqrt{2}} \Delta_{0\Uparrow} ({\bf k}), 
\nonumber \\
{\widetilde \Delta}_{10} ({\bf k}) 
& = &
\frac{1}{\sqrt{2}} \Delta_{\Uparrow \Downarrow} ({\bf k})
-
\frac{1}{\sqrt{2}} \Delta_{\Downarrow\Uparrow} ({\bf k}), \\
{\widetilde \Delta}_{1{\bar 1}} ({\bf k}) 
& = &
\frac{1}{\sqrt{2}} \Delta_{0\Downarrow} ({\bf k})
-
\frac{1}{\sqrt{2}} \Delta_{\Downarrow 0} ({\bf k}),
\nonumber
\end{eqnarray}
for the triplet sector $S = 1$, 
with $m_s = \{+1, 0, -1 \} = \{1, 0, {\bar 1} \}$,
and 
\begin{eqnarray}
{\widetilde \Delta}_{22} ({\bf k}), 
& = &
\Delta_{\Uparrow \Uparrow} ({\bf k})
\nonumber \\
{\widetilde \Delta}_{21} ({\bf k}) 
& = &
\frac{1}{\sqrt{2}} \Delta_{\Uparrow 0} ({\bf k})
+
\frac{1}{\sqrt{2}} \Delta_{0\Uparrow} ({\bf k}), 
\nonumber \\
{\widetilde \Delta}_{20} ({\bf k}) 
& = &
\frac{1}{\sqrt{6}} \Delta_{\Uparrow\Downarrow} ({\bf k})
+
\sqrt{\frac{2}{3}} \Delta_{0 0} ({\bf k})
\frac{1}{\sqrt{6}} \Delta_{\Downarrow\Uparrow} ({\bf k}), \\
{\widetilde \Delta}_{2{\bar 1}} ({\bf k}) 
& = &
\frac{1}{\sqrt{2}} \Delta_{0\Downarrow} ({\bf k})
+
\frac{1}{\sqrt{2}} \Delta_{\Downarrow 0} ({\bf k}), 
\nonumber \\
{\widetilde \Delta}_{2{\bar 2}} ({\bf k}) 
& = &
\Delta_{\Downarrow \Downarrow} ({\bf k}),
\nonumber 
\end{eqnarray}
for the quintuplet sector $S = 2$, 
with $m_s = \{+2, +1, 0, -1, -2 \} = \{2, 1, 0, {\bar 1}, {\bar 2} \}$.

This concludes the supplementary material provided for the study of 
unconventional color superfluid phases of three-color fermions in the 
presence of color-orbit, color-flip and tunable interactions in a single 
s-wave channel. The physics described here may be investigated in fermionic
isotopes of Lithium, Potassium and Yterbium, when three internal states 
are trapped and interactions are adjusted via Fano-Feshbach resonances.

\end{document}